\documentclass{aa}
\usepackage{graphicx}
\begin{document}

%

%
   \title{The coronal FeXXI $\lambda$1354.094 line in\\ AB Doradus
\thanks{Based on observations with the NASA/ESA {\it Hubble Space telescope},
obtained at the Space Telescope Science Institute, which is operated by the
Association of Universities for Research in Astronomy, Inc., under the NASA 
contract NAS 5-26555.}  }

  \author{O. Vilhu
\thanks{{\it Hubble Space Telescope} Guest Observer}
         \inst{1}
       \and
        P. Muhli \inst{1}
        \and
          R. Mewe \inst{2}
         \and
          P. Hakala \inst{3}
          }

   \offprints{O. Vilhu}

   \institute{
Observatory, Box 14, FIN-00014 University of Helsinki. Finland\\
      \email{osmi.vilhu@helsinki.fi, muhli@astro.helsinki.fi} 
\and 
   SRON Laboratory for Space Research, Sorbonnelaan 2,
               3584 CA Utrecht, The Netherlands\\
              \email{R.Mewe@sron.nl}
\and
   Tuorla Observatory, University of Turku, Finland\\
          \email{pahakala@astro.utu.fi}
                                           }

   \date{Received  ; accepted  }


\abstract{
The active late-type star AB Doradus was observed in February 1996
with the Goddard High Resolution Spectrograph of the {\it Hubble Space Telescope} using the low resolution G140L grating. The observations covered one half
of the star's rotation cycle (P = 0.514 d) with 11.5 min time resolution.
The strong coronal Fe XXI $\lambda$1354.094 line formed at 10$^7$ K was analysed
and its emission measure (EM) derived.  This EM  is much higher than that 
derived from  recent XMM-Newton observations  (G\"udel et al. 2001), and
earlier EXOSAT (Collier Cameron et al. 1988) and ASCA/EUVE  (Mewe et al. 1996)
data, as well,  requiring a variability by a factor of 5. The physical reason
for the variability remains unknown, since (outside flares) the observed
broad band  variability of AB Dor is much smaller.
\keywords{ stars:coronae -- stars: activity -- stars: individual: AB Dor
                     -- ultraviolet: stars }
    }

        \maketitle

\section{Introduction}

AB Doradus is a young and rapidly rotating  late-type star (K1 IV, P$_{rot}$ =
0.514 d, vsin{\it i} = 100 km/s) whose corona has been extensively studied due 
to its brightness and
activity. The hot (10$^{6-7}$ K) corona of AB Dor is best visible in the 
extreme ultraviolet
(Rucinski et al. 1995, EUVE) and soft X-rays (Collier Cameron et al. 1988,
EXOSAT; Vilhu et al. 1993, GINGA; Mewe et al. 1996, ASCA;
K\"urster at al. 1997, ROSAT; G\"udel et al. 2001, XMM-Newton). In particular, 
XMM-Newton confirmed the low coronal iron abundance found by the EUVE/ASCA-
combination (Mewe et al. 1996), which was 4-5 times  lower than  the  
photospheric abundance  (Vilhu et al. 1987). Further, XMM-Newton, EUVE and 
ASCA have permitted 
very detailed studies of the coronal temperature stratification 
using lines formed at different temperatures.

A few weak coronal lines also exist at longer wavelengths. A potentially
important case is FeXXI $\lambda$1354.094 which is an M1-type forbidden 
transition between two fine structure
levels of the ground state of FeXXI. The line is 
optimally formed at 10$^7$ K and is
accessible to the Hubble Space Telescope (HST) spectrographs. Using HST, the line 
was discovered in HR 1099 (Robinson et al. 1996), Capella (Linsky et al. 1998)
and AU Mic (Pagano et al. 2000) giving an opportunity to study
the whole chromospheric-coronal complex simultaneously using  far
ultraviolet lines only.  

Using HST, we observed the FeXXI-line of AB Dor during half of its 
rotational  period. We present the results and  methods
to analyse the line strength in terms of the emission measure.      
 
\section{ Observations}

  The observations were performed with the {\it Goddard High Resolution
Spectrograph} (GHRS) onboard the {\it Hubble Space Telescope} (HST) on
February 5th, 1996. This was a repeat run of our original programme
(ID 5310) which failed partially in November 1994. The target was in
the Continuous Viewing Zone (CVZ), 
hence we were able to acquire a continuous series of spectra
interrupted only by three South Atlantic Anomaly (SAA) passages. We
used the G140L low-resolution grating which
provides a resolving power $R = \lambda / \delta \lambda
\approx 2100$ at the coronal FeXXI 1354 \AA ~line, giving a spectral
resolution of $\sim 150$ km s$^{-1}$. The spectra which covered
a wavelength range of $\sim 1290-1580$ \AA ~were captured with the D1
Digicon detector using the 2.0 arcsec square Large Science Aperture
(LSA). The science exposures were preceded by a Spectrum Y Balance (SPYBAL) 
calibration lamp exposure with a slightly different carrousel position
for the G140L grating as compared to the subsequent target
observations. However, we made use of the calibration spectrum to determine
the zero-point offset of the default wavelength solution of the spectra,
providing an adequately accurate wavelength calibration as to the low
spectral resolution. 

The science exposures, obtained at 03.796-09.943 (heliocentric) UT
(about one half of the rotation period of our target) using the ACCUM
mode, resulted in 32 spectra with 11.53 min average time resolution
(including overheads and SAA passages). The default COMB=FOUR setting was 
used to compensate for diode response variations. The response
irregularities due to photocathode granularity were not corrected for
(FP-SPLIT=NO) and only the half-diode width subsampling method
(STEP-PATT=4) was implemented, since we desired to maximise the time
resolution of our observations.      
 
The spectra were reduced with the IRAF (Image Reduction and Analysis Facility)
and STSDAS (Space Telescope Science Data Analysis System)
software. Utilising the best calibration
reference files available at the time of the reductions (February
1998) we used the {\it calhrs} calibration 
routine to assign wavelength solution, flux values and
error estimates to the raw GHRS data.
Furthermore, the wavelength scale of the science
spectra was adjusted using the {\it waveoff} task and the SPYBAL
calibration spectrum so that any zero-point offset from the default
wavelength solution could be corrected for.

 Fig. 1 shows the average spectrum
 excluding the flare spectrum at phase 0.8 (see Fig.3).
The mean intensity (1.25 $\pm{0.05}$) 10$^{-12}$ erg cm$^{-2}$ s$^{-1}$  
  of the strongest line, the CIV 1549 doublet, is very close to  that observed  
one year earlier with HST (Vilhu et al. 1998; Brandt et al. 2001). 
  
Fig. 2 shows an enlargement of the FeXXI $\lambda$1354.094 region. 
For comparison, the solar-quiet network spectrum (multiplied by 10
and scaled to  AB Dor's distance of 15 pc) as observed
by SUMER on board SOHO (Curdt 1998, Wilhelm et al. 1999) is overplotted. 
The solar spectrum, broadened by the instrumental (150 km/s) and 
rotational (100 km/s) profiles, is also shown. 

Fig. 3 shows  light curves  
of selected lines  (marked in Fig. 1) using the canonical ephemeris (Innis et al. 1988):
 $$phase(0.0) = HJD 2444296.575 + 0.51479E . \eqno(1)$$

The flare at phase 0.8 did not produce apparently any hot optically thin gas
since no trace of it is visible in the Fe XXI 1354.094 line.

%
   \begin{figure*}
\resizebox{12cm}{!}{\includegraphics{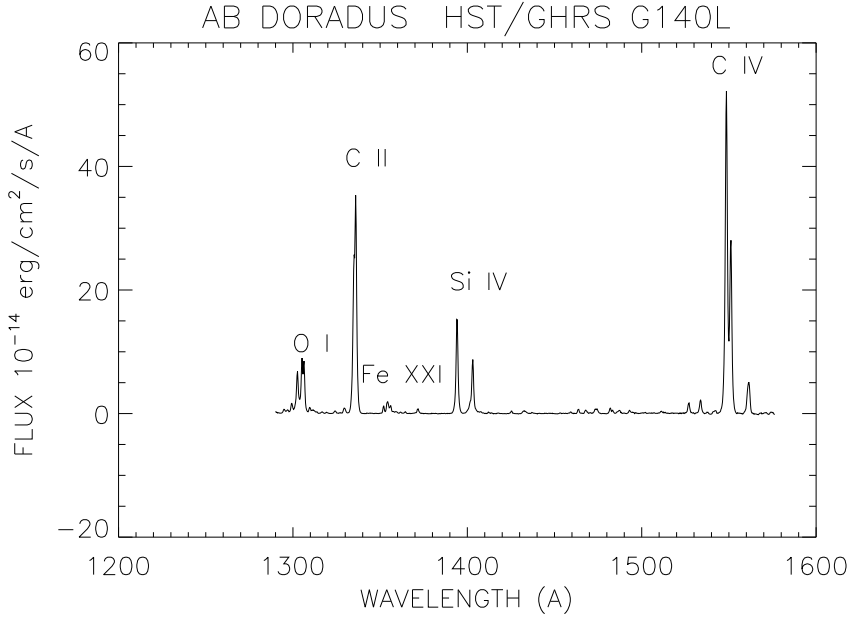}}
\hfill      
\parbox[b]{55mm}{\caption{ The spectrum of AB Doradus as observed with the G140L 
grating of the Goddard High Resolution Spectrograph of the Hubble Space Telescope.}}
\label{FigSpec}
\end{figure*}
   \begin{figure*}
\resizebox{12cm}{!}{\includegraphics{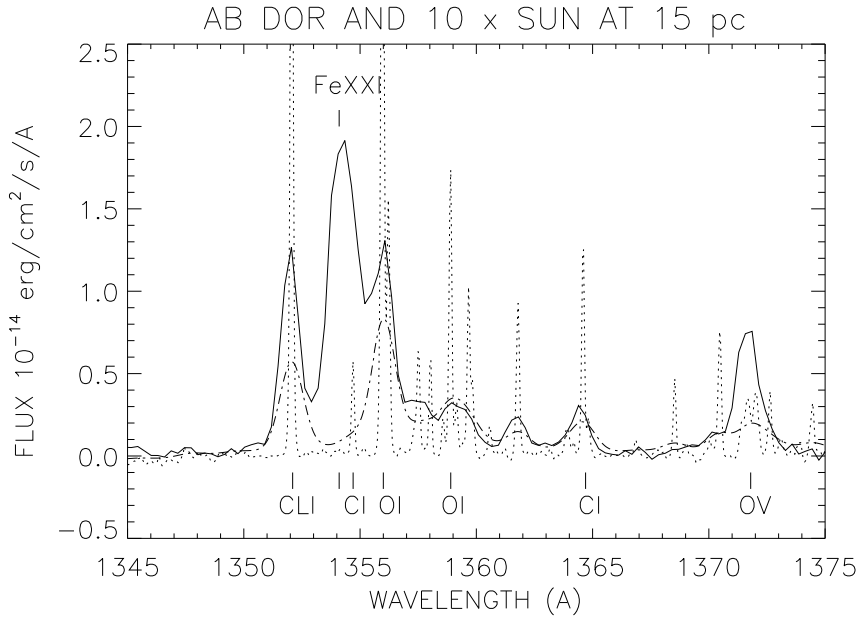}}
\hfill      
\parbox[b]{55mm}{\caption{ An enlargement of the spectrun in Fig. 1 around 
the FeXXI $\lambda$1354.094 line (the solid line).  
A quiet network solar spectrum multiplied by 10 and 
scaled to a distance of 15 pc is also shown with and without 
instrumental (150 km/s) and rotational (90 km/s) broadening 
(dash-dot and dashed lines).}}
\label{FigZoom}
\end{figure*}
%
   \begin{figure*}
\resizebox{12cm}{!}{\includegraphics{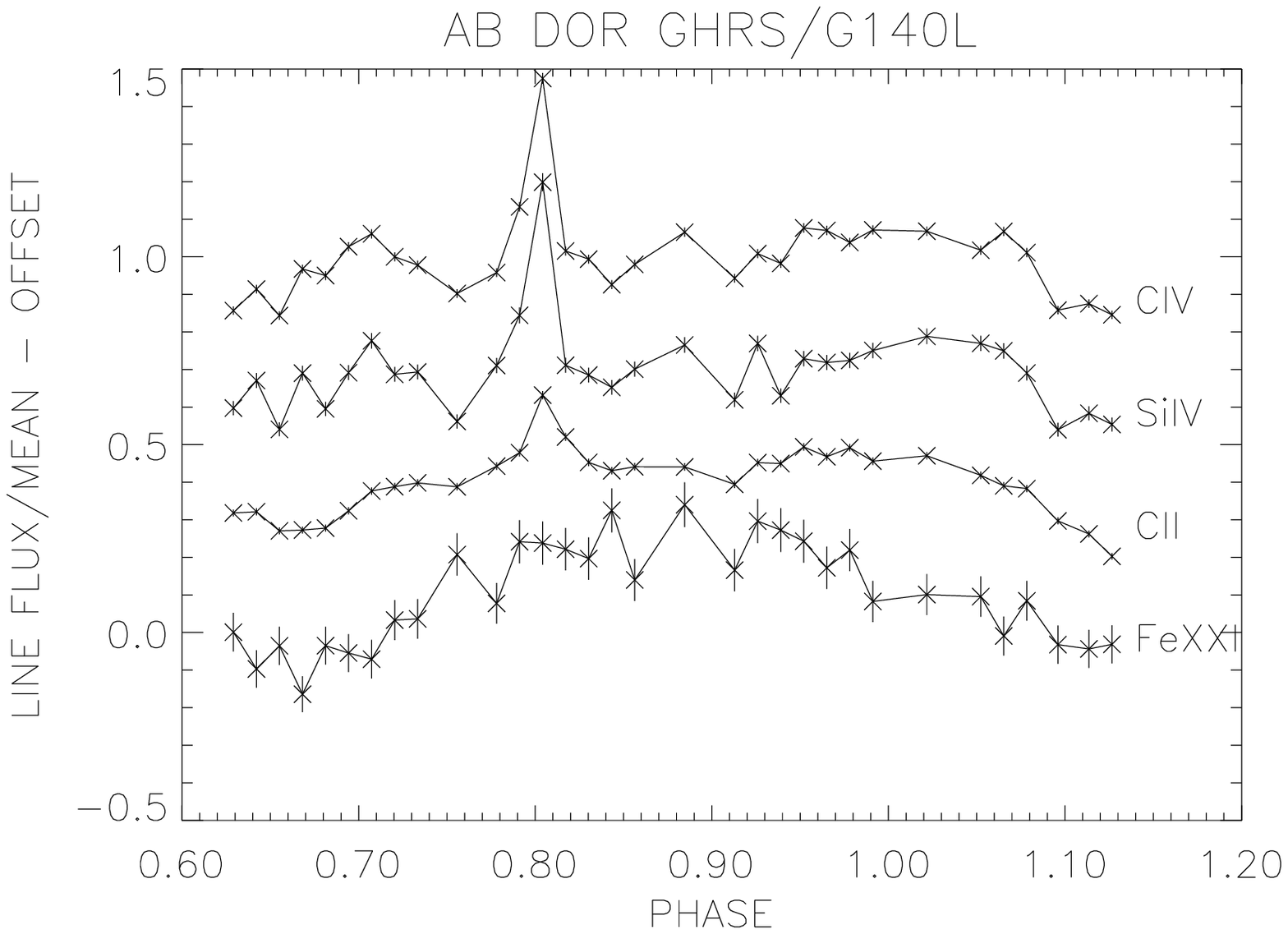}}
\hfill      
\parbox[b]{55mm}{\caption{ Light curves of total intensities of selected lines 
marked in Fig. 1. The intensities are scaled with their mean values and shifted by
0, -0.3, -0.6 and -0.9 for CIV, SiIV, CII and FeXXI, respectively. }}
\label{FigZoom}
\end{figure*}

\section{   Calculation of the FeXXI $\lambda$1354.094 line strength}        

A multi-Gaussian fit was applied to the region of the spectrum shown in Fig. 2 resulting
in the line parameters given in Table 1.

The coronal  FeXXI 1354.094 line (Dere et al. 1997) 
is a blend with  CI 1354.288 and  cannot be separated with our
resolution. Fortunately, there is another  CI line nearby at 1364.164  which is
stronger in the Sun (see the dotted line in Fig.~2) and Capella (Linsky et al. 1998)  
compared to the $\lambda$1354.288 line.
Hence, we can safely assume that its contribution is also small  in AB Dor. 
The blending with the nearby
OI 1355.60 line can be resolved with  Gaussian fits and the FeXXI line flux 
(3.03 $\pm{0.3}$ 10$^{-14}$ erg cm$^{-2}$)  properly estimated.

We calculate the line strength of the Fe XXI $\lambda$1354.094 line
which
is an M1-type forbidden transition between two (of four total) fine structure levels (1-2) of
the ground state $^3$P. We follow the procedure given by Mewe et al. (1985).

The line strength (photons/cm$^3$/s) of a given transition from level $j$ to
level
$k$ (not necessarily the original level $i$ from which the line was
excited) is
given by
$$P_{jk}^{exc} = S_{exc}\ BR\ n_e\ N_{ion}, \eqno(2)$$
where $BR \equiv A_{jk}/\sum_{\ell\le j} A_{j\ell}$ is the branching
ratio (the $A$'s are radiative transition probabilities),
n$_e$ is the electron density (in cm$^{-3}$),
$N_{ion}$ is the concentration (in cm$^{-3}$) of the radiating ion,
and
$S_{exc}$ is the electron collisional
excitation rate coefficient (in cm$^3$~s$^{-1}$)
given by
$$S_{exc} = 8.63\ 10^{-6} \left(\bar \Omega(y) \over w_i\right) T^{-
1/2} e^{-y},
\eqno(3)$$
where $y = E_{exc}/kT = 1.43877\ 10^8\ a/ \lambda T$ (electron
temperature
$T$ in K and wavelength $\lambda$ in \AA). Here $a=E_{exc}/h\nu$, the
ratio
of line excitation and photon energy ($a > 1$ for a line transition
not ending
on the ground level).
Further, $\bar \Omega(y)$ is the collision strength averaged over a
Maxwellian
electron energy distribution  and $w_i$ is the statistical weight of the
initial
lower (usually ground) level.

The reduced line strength $P'$ in energy units of 10$^{-23}$ erg
cm$^3$ s$^{-1}$ is given by
$$P' \equiv 10^{-Q} 
\equiv 10^{23} P/(n_e N_H),  \eqno(4a)$$

or
 
$$P' =        10^{23} BR\ E_{ph}\ S_{exc}\ (N_{ion}/N_{el})(N_{el}/N_H),\eqno(4b)$$

where
$E_{ph}$ is the photon energy in ergs (= 1.9863 10$^{-8}$/$\lambda$(\AA)),
$N_{ion}/N_{el}$ is the ion fraction given by Arnaud and Rothenflug (1985, Ar-Ro)
or Arnaud and Raymond (1992, Ar-Ra),
$N_{el}/N_H$ is the iron abundance relative to Hydrogen
and $Q$ is the exponent given in the line flux table IV of Mewe et al. (1985).
We adopt 4.68 10$^{-5}$  as the solar iron abundance relative to 
hydrogen (Anders \& Grevesse, 1989).

Substituting relevant data and rewriting, in this special, gives:
$$P' = 253.7(Fe/Fe_{\sun})\ {{BR\bar \Omega(y)}\over {\lambda w_i}}
{N_{ion}\over N_{el}} {1\over \sqrt{T}}  e^{-14.3877 a/\lambda T},\eqno(5)$$

where $T$ is in units of 10$^7$~K, $\lambda$ in \AA\ and Fe/Fe$_{\sun}$ is
the coronal iron abundance relative to the solar photospheric one.
For transitions from the
ground state
$a$=1 and $w_i$=1 and for the $\lambda$1354 line we also have $BR$=1.
The collision strength $\bar \Omega(y)$ is taken from the calculations by Aggarwal (1991) who takes into account the effect of
near-Threshold resonances.

Collisional depopulation of level 2 begins at
an electron density ($n_e$)
where $n_e S_{exc}$ exceeds  A, i.e., at n$_e$  $\sim$~4~10$^{14}$ cm$^{-3}$,
well above typical stellar coronal densities and we neglect these effects.

However, the excitations to neighbouring levels 1-3 and 1-4 also contribute
nearly
fully to the population of the upper level 2 of the 1354 line because
the
radiations from  levels 3 and 4 ultimately cascade down to level 2. With the collision strengths from Aggarwal (1991) and branching ratios based on transition probabilities from the CHIANTI data base
(Dere et al. 1997) we estimate that the cascades enhance the excitation rate 1-2 by a factor of 
 $F \simeq$~2-2.2 which gives 

  $$P' \equiv 10^{-23} (Fe/Fe_{\sun}) 10^{-Q^{Eff}}{\it erg cm^3 s^{-1}}, \eqno(6)$$.

Table 2 gives  values of $Q^{Eff}$ and some other line parameters for
three  temperature-values. Using a denser temperature-grid,  
a very  good numerical fit was obtained by 
Q$^{Eff}$ = constant + 16.6$\times$(log$_{10}$T - 6.96)$^2$. 

The contribution to the observed line flux  from a plasma volume $V$ with temperature {\it T} (in K) and emission measure $EM = n_e N_H V$
can then be computed by

 $$f = EM \times P'/(4\pi d^2), \eqno(7)$$
where the distance $d$ of AB Dor is 15 pc (see  Vilhu et al. 1998 and G\"udel et al. 2001).

Using the observed Fe XXI flux value from Table 1 and assuming  
solar iron abundance,
this formula gives a value (3.75 - 4.50)$10^{52}$ cm$^{-3}$ 
for the emission measure at 10$^7$ K depending on the ionisation fraction model (Ar-Ra or
Ar-Ro).
This value is in fact  the {\it lower} limit if other temperatures contribute as well.

A multitemperature fit to the combined EUVE and ASCA data by Mewe et. al. (1996) and 3-T
fits to the XMM-Newton RGS2 and pn-CCD data (G\"udel et al. 2001) give very similar total
emission measures when integrated over temperature bins ( (7.5 - 9.5) 10$^{52}$ cm$^{-2}$) and
a low iron abundance  $Fe/Fe_{\sun}$ = 0.22.  If such a low iron abundance is used to compute
the emission measure for the Fe XXI 1354 line, the resulting EM  would be much
higher ((17 - 20) 10$^{52}$ cm$^{-2}$). 
Using a  Bremsstrahlung continuum model to fit the EXOSAT (LE + ME) spectrum, 
Collier Cameron et al. (1988) arrived at a  value of (7 - 10) 10$^{52}$ cm$^{-2}$ 
(scaling to a  distance of 15 pc). All these values are plotted in Fig. 4 where 0.3 dex
is assumed for the 1-T and  3-T fits of HST, XMM and EXOSAT while a  more dense (0.1 dex)
binning was used for the ASCA/EUVE data.

   \begin{figure*}
\resizebox{12cm}{!}{\includegraphics{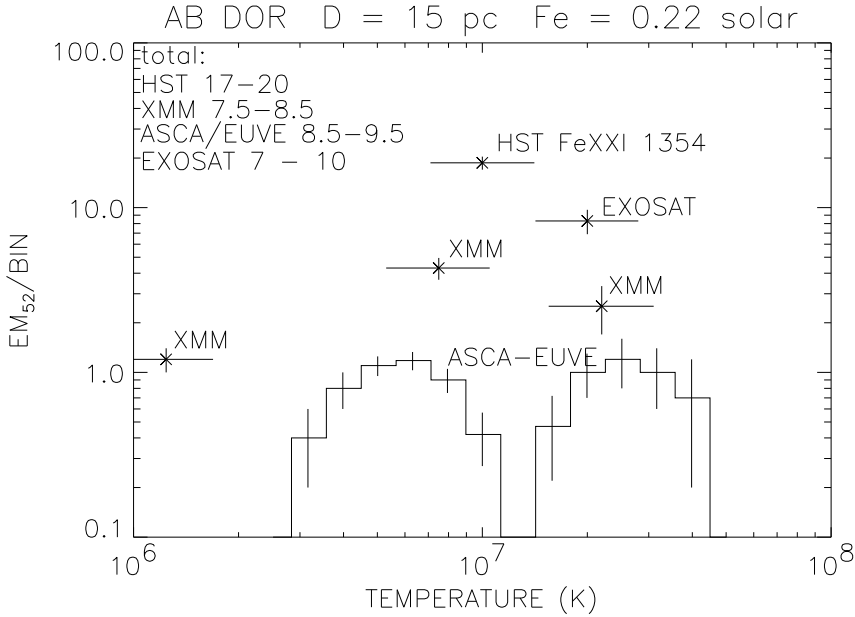}}
\hfill      
\parbox[b]{55mm}{\caption{ Emission measures vs temperature (in units of 
10$^{52}$ cm$^{-3}$ bin$^{-1}$, where bin = the temperature range of the emission
measure in question) from different missions and observing periods 
are collected and compared. The total emission measures (summed over all temperature bins)
are given in the upper-left corner.   
                                }                     }
\label{EM}
\end{figure*}

\section{Discussion and Conclusions}

A striking feature of Fig. 4 is the similarity of the total emission
measures obtained from different missions (EXOSAT, ASCA/EUVE, XMM-Newton) with
different observing times, energy bands and spectral resolutions. 
The emission measure derived in the present paper from the Fe XXI 1354.094 line
deviates significantly from these, especially when the shallow
line formation temperature range is considered.  

Large variability would,
of course, explain the discrepancy. During the present observations
the variability was $\pm{15}$ per cent and the 5.5 years of ROSAT monitoring
did not resolve any pronounced variability in the soft X-ray flux
(K\"urster et al.  1997). 
However, we cannot exclude large variabilities in the shape of
the differential emission measure curve, especially in the hot region
where the Fe XXI line is formed. 
 By comparing the FeXXI 1354 line
with  other FeXXI lines in the EUVE-range,  
Linsky et al. (1998) concluded that there was 
large variability in Capella at 10$^7$ K.
Using the {\it HST/STIS} Johnson et al. (2001) 
also concluded a
large (factor of 5) variation of the FeXXI 1354 line in the  G8 component of Capella. 
Hence, the variability may be the reason also
to cause the unusual strength of the line in AB Dor, as well,
but the parameter responsible for variations remains unsolved.

An explanation would be if the abundance at 10$^7$ K
is higher  and closer to the photospheric value (and possibly also variable), despite the 
ASCA/EUVE and XMM-Newton spectroscopic results.  This would, however,
require a rediscussion of these data  to see whether such
an abundance gradient  is feasible.
The problem is related to the inverse FIP-effect found by XMM-Newton 
in AB Dor (G\"udel et al. 2001) and in HR1099 (Brinkman et al. 2001).
If the low-FIP elements (like iron)  are deficient  in the  corona
a natural  question is 'where  have they gone?.
 One possibility is that they have been
accelerated to the top of the corona (coronal loop apexes above 10$^7$ K) 
where the
abundances are consequently larger. If such a temperature gradient is introduced to the fitting of XMM-Newton data the result would probably be a 
satisfactory Fe XXI 1354 line but  a too strong
iron K$\alpha$ line instead.

Marc Bos has been observing AB Dor frequently at his Mt Molehill observatory.
 The optical light curve between Sept 8, 1995 -- Jan 8, 1996 showed a
broad minimum between phases 0.2 -- 0.6 (Bos 2000). Hence, although the HST observations
were slightly outside this range (Feb 5, 1996), it is probable that we
observed AB Dor with HST during its maximum light, i.e. when the 
photospheric spot complexes
were situated 
behind the limb. If the small trend seen in Fig. 3 (broad maximum
in the Fe XXI 1354 light curve) is interpreted as due
to the rotational modulation, this would mean that large  active region 
 and spot complexes  were situated in  opposite hemispheres.

During a flare G\"udel et al.
(2001) found the iron abundance to rise by a factor of 3. This could also 
partially
solve the EM-problem discussed above,  if the trend in Fig. 3 was  part of a long-lasting flare 
(around the chromospheric-transition region flare visible at phase 0.8, see Fig. 3).

Although the spectral resolution of the G140L grating (resolving power = 2000)
does not permit a detailed discussion of the line profiles, a few remarks
concerning the line widths in Table 1 are worthwhile. All the lines are somewhat
redshifted above the radial velocity of the star +30 km/s. The widths of the
narrowest lines (Cl 1351.66 and OI 1355.60) are similar to those of
Si IV 1400 and CIV 1549 (210 -- 220 km/s). The width of this size can be
explained by the combined rotational (vsin{\it i} = 100 km/s) and instrumental
(150 km/s) profiles. The larger width of the Fe XXI line (325 km/s) cannot
 be explained by  larger thermal velocity (90 km/s at 10$^7$ K) alone.
It requires either  larger rotational broadening 
or else an additional turbulent component of size 110 km/s (or both). Large
rotation  would be natural if an extended co-rotating corona
(loops) is involved from the top of which the Fe XXI line originates.

\begin{table}
\caption{\label{tab:line}\protect\small Observed parameters of selected lines in the FeXXI 1354 region (Fig.2). The line flux is in units of  
 10$^{-14}$ erg cm$^{-2}$ s$^{-1}$ and FWHM in units of  km s$^{-1}$. }
\small
\begin{center}
\begin{tabular}{ccccc}
\multicolumn{5}{c}{ } \\ \hline

$\lambda$ (obs)  &  flux  & FWHM & Identification   &                    \\
\hline

1351.99 & 1.25 $\pm{0.09}$ & 210 $\pm{20}$    &  Cl I 1351.66 &  \\

1354.26     &  3.03 $\pm{0.3}$ & 325 $\pm{40}$    &  Fe XXI 1354.094 &   \\

1355.98     &  1.18 $\pm{0.1}$ &  225 $\pm{30}$&      O I 1355.60 &                     \\

2364.39   &   0.34 $\pm{0.05}$ &  255 $\pm{30}$  &    C I 1364.16  &                 \\

1371.66  &    1.04 $\pm{0.10}$ &  285 $\pm{40}$  &   O V 1371.29  &                       \\

\hline
\end{tabular}
\end{center}
\end{table}


\begin{table}
\caption{\label{tab:line}\protect\small FeXXI $\lambda$1354.094  line parameters.  The two values of N$_{FeXXI}$/N$_{Fe}$ and 
  Q$^{Eff}$ refer to the Ar-Ra (upper) and Ar-Ro (lower) 
ionization models.  }            
\small
\begin{center}
\begin{tabular}{ccccc}
\multicolumn{5}{c}{ } \\ 

logT  & 6.8 &       7.0 &       7.2   &                    \\

N$_{FeXXI}$/N$_{Fe}$ &  0.0832 & 0.275     &   0.0214 &  \\

     - '' -          &  0.0617 & 0.204     &   0.0158 &   \\

$\bar \Omega(y)$     &  2.627e-2 &  1.990e-2 &  1.493e-2 &                     \\

Q                    &  3.525  &  3.073   &   4.358  &                 \\

F                   &   2.016 &  2.134   &   2.236  &                       \\

Q$^{Eff}$  &  3.090  &  2.665   &     3.876 &    \\
         
     - '' -  & 3.220       & 2.744      &    4.008  &     \\

\end{tabular}
\end{center}
\end{table}
\begin{acknowledgements}
This work was performed with support from the Academy of Finland (OV)
and Space Sesearch Organization of Netherlands (SRON) which is supported
financially by NWO (RM). We are grateful to  Diana Hannikainen for reading
the manuscript and correcting our crude language.  We thank Werner Curdt 
for providing the solar SUMER-spectrum and Marc Bos for information on his
 optical photometry.

\end{acknowledgements}

\end{document}